\documentclass[fleqn,10pt,twocolumn]{wlscirep}
\usepackage[utf8]{inputenc}
\usepackage{mathpazo}
\usepackage[T1]{fontenc}
\usepackage{bm}
\usepackage{amsmath}
\usepackage{amssymb}
\usepackage{graphics}
\usepackage{graphicx}
\usepackage{float}
\usepackage{subfigure}
\usepackage{color}
\usepackage{hyperref}
\usepackage{amsmath}
\usepackage{textcase}
\usepackage{txfonts}
\usepackage{tabularx}
\usepackage{array}
\usepackage[switch]{lineno} 
\usepackage{indentfirst}

\title{Dirac points and Weyl phase in a honeycomb altermagnet}

\author[ ]{Meng-Han Zhang}
\author[ ]{Xuan Guo}
\author[*]{Dao-Xin Yao}
\affil[ ] {Guangdong Provincial Key Laboratory of Magnetoelectric Physics and Devices, State Key Laboratory of Optoelectronic Materials and Technologies, Center for Neutron Science and Technology, School of Physics, Sun Yat-Sen University, Guangzhou, 510275, China}
\affil[*]{yaodaox@mail.sysu.edu.cn}

\begin{abstract}
We present unconventional nodal crossings in a two-dimensional (2D) collinear altermagnet, which are enforced by crystal symmetries to lock spin polarization and valley degrees of freedom. The altermagnetism generate nonrelativistic spin-splitting in honeycomb antiferromagnets, guaranteeing novel band degeneracies between bands sharing identical spin configurations yet different sublattices. Inspired by the $XPS_{3}$ (X=Mn, Fe, Ni) materials, we demonstrate distinctive Berry curvature distributions concentrating intensely at Weyl nodes, which further generalize the locking between valleys and Berry curvature. Topological phase transitions are characterized by the high Chern numbers preserving the non-intersecting flows of Wannier centers over occupied bands, where degeneracy lifting contributes to unconventional spin textures to induce the valley Hall effect. Our results yield unique topological nodes via leveraging the crystal symmetry constraints with the intrinsic time-reversal symmetry breaking, where corresponding topological responses enable the potential of advancing spintronics. Our results yield unique topological nodes without the spin-orbit coupling (SOC) achieved by combining crystal symmetry constraints with inherent time-reversal symmetry breaking, whose associated topological responses enable promising applications in spintronics.
\end{abstract}

\begin{document}
\maketitle

\thispagestyle{empty}

\begin{figure}[ht]
\centering
\includegraphics[width=3.5 in]{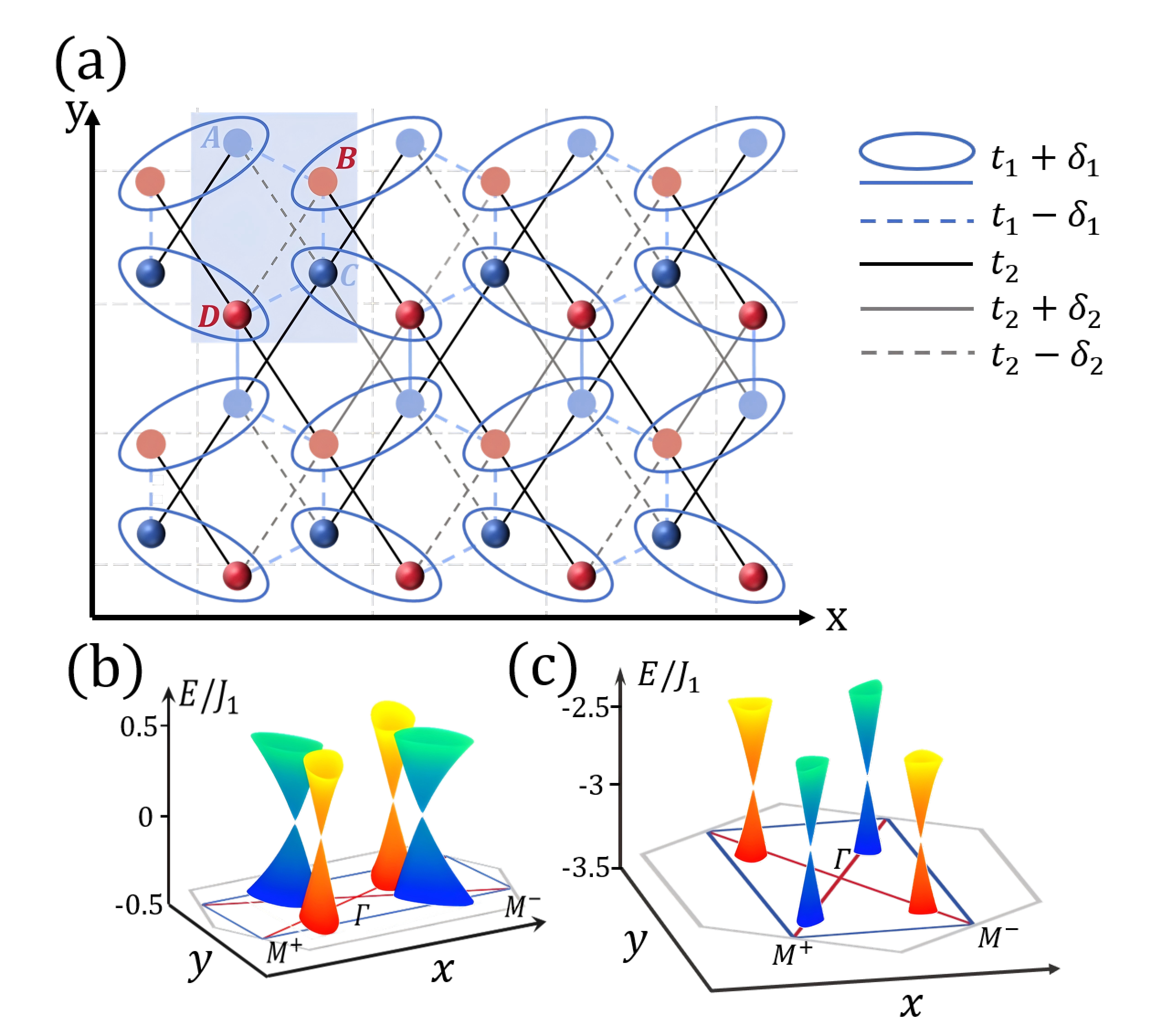}
\caption{(a) Schematic of the honeycomb-lattice model for altermagnetism in group p6m. The model has four sublattices at each lattice point with red and blue dots denoting anti-aligned moments. The blue dashed lines represent the nearest-neighbor (NN) hopping strengths $t_{1}$-$\delta_{1}$, and ellipses accompained with blue lines are hopping terms with negative signs $t_{1}$+$\delta_{1}$. The crystalline environment generates anisotropic second-nearest-neighbor hopping (black solid and dashed lines) $t_{2}$ and $\delta_{2}$. (b) Dirac crossings with spin-up and spin-down bands develop on the BZ edges. (c) Weyl points between bands of the same spin are shown in red and blue, respectively. The massless crossings along the high-symmetry directions $\Gamma-M$ annihilate each other into quadratic crossings with bands of opposite mirror $M_{x}$ eigenvalues $\pm$1.}
\label{fig1}
\end{figure}

Unconventional Zeeman splitting can induce diverse nontrivial topology in two-dimensional (2D) collinear antiferromagnets, while preserving symmetry-constrained zero net magnetization\cite{SpinSplitting1,Landscape4}. We investigate a honeycomb lattice possessing combined time-reversal ($\mathcal{T}$) and mirror symmetries ($\mathcal{M}$), drawing partial inspiration from $XPS_{3}$ compounds (X=Mn, Fe, Ni)\cite{AIMaterials,RPPPAMaterials,TwoDimensionalMaterials,AltermagneticMetal,MultifunctionalAntiferromagneticMaterials}, where the altermagnetism enforces precise compensation of momentum-dependent spin polarizations in the absence of spin-orbit coupling (SOC)\cite{SpinSplitting6,SpinSplitting8}. Pairwise topological nodes are protected by altermagnetic symmetry due to the breaking of $\mathcal{PT}$ symmetry, which enable highly localized Berry curvatures that underlie the emergence of the valley Hall effect\cite{ValleyTransport1,ValleyTransport2}. Although the net Berry curvature cancels out entirely as the valley-resolved contributions exhibit identical magnitudes yet opposite signs among the four valleys\cite{Graphene1,SpinSplitting7,RotationSymmetry}, we demonstrate a finite net Berry curvature by reconstructing altermagnetic order achieved via staggered spin configurations and modulated coupling strength. The formation of gapless chiral edge states and dissipationless spin currents is imposed by lifting the band degeneracies at a general position, which manifest observable transport phenomena\cite{NonlinearTransport}.

We study the spin current generation induced by momentum-dependent spin splitting rather than the relativistic SOC, realizing emergent topological phases via four-sublattice bond dimerization\cite{dimerization1,dimerization2}. We demonstrate Dirac cones and pairwise Weyl nodes by focusing on the collinear antiferromagnetic order and crystal symmetry\cite{WeylAltermagnetism,WeylFermions,WeylPoints}, which can be categorized into $d_{x^{2}-y^{2}}$ or $i$-wave altermagnet in the honeycomb monolayer\cite{dWaveAltermagnets,BandSplitting4}. Symmetry-protected Weyl points move along high-symmetry paths in the first Brillouin zone (BZ), ultimately merging and annihilating each other via the dimerization of hopping terms\cite{SpinSplitting4,SpinSplitting5}. Although SOC is considered negligible, it enables significant topological responses rather than incompatible with altermagnetism, including the anomalous Hall effect (AHE) in conjunction with emergent weak ferromagnetism. We calculate the Hall conductivity grounded in a tight-binding model formed by hexagonal clusters, where the behaviors of valley-polarized currents are imposed by the gapped nodes in altermagnetic structure\cite{ValleyEdgeInsulator}. When integrated with proximity-induced superconductivity, the pairing symmetries and finite-momentum pairing in altermagnets offer a fertile ground for researching novel topological phases and superconducting behaviors\cite{ZhangSongBo1,zhudi1}. 

In this work, we demonstrate the existence of band crossings in 2D honeycomb lattice, including Dirac and Weyl nodes, where their topological responses under SOC are determined by the altermagnetic order. The spectrum of nanoribbons supports localized edge states that span the whole zigzag terminations with pronounced van Hove singularities (vHSs), while the armchair configurations exhibit more uniform band structures rather than sharp spikes in the density of states (DOS)\cite{VanHoveSingularities}. We align two ribbons with complementary bond patterns to ensure effective overlap of the edge states, promoting efficient state mobility dynamically via geometric or bond modulation\cite{Graphene2,SpinGroup5,SpinGroup6}. The strong-weak bonds alter the quantum interference between electron wavefunctions, contributing to the localization of edge states for quenching kinetic energy. Our research elucidates the intricate interplay between altermagnetism and topological phases, providing a promising platform for highly efficient spintronics devices that are not confined to heavy-element materials\cite{SpinSplitting2,SpinSplitting3}.

\begin{figure*}[ht]
\centering
\includegraphics[width=6.7 in]{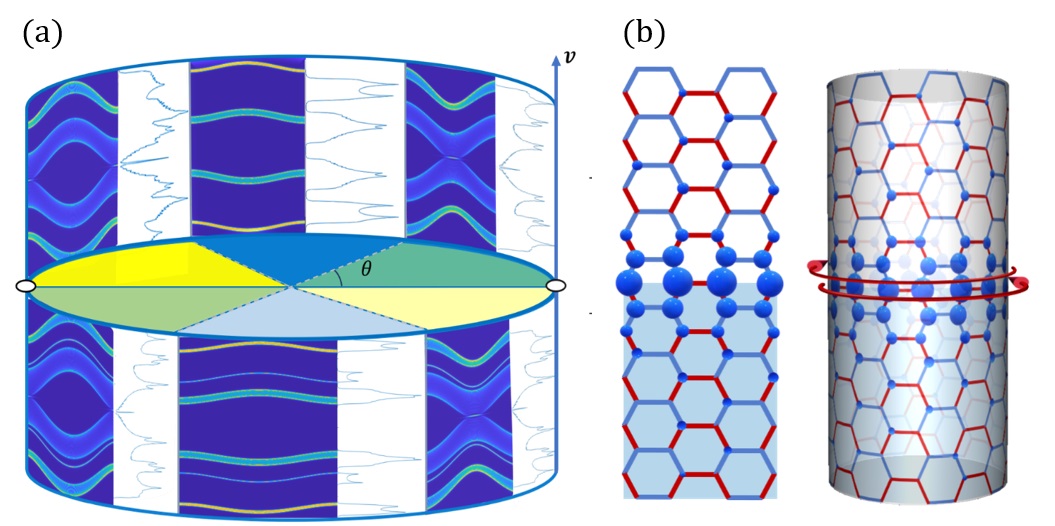}
\caption{(a) Analytical demonstration of the phase diagram formed by $\nu$ and $\theta$. The Dirac points merge together at the phase boundaries $\theta$ = $\pm\arctan\frac{\sqrt{5}}{2}$ and $\pm(\pi-\arctan\frac{\sqrt{5}}{2})$, where the spectral function of edge modes and DOS are depicted for each separated phase. (b) Two pairs of winding numbers with even and odd parity. The parameters of $\nu$ are chosen at equal intervals from -1.2 to 1.2. (c) Armchair edge modes periodic along $k_{1}$ are regarded as a ribbon on a cylinder, where the amplitude of wave function is localized in the interface. (d) The energy spectra correspond to the propagating edge modes with double degeneracy, which indicate $(1, -1)$ for $\nu$ $>$ 0 and $(0, 0)$ for $\nu$ $<$0.}
\label{fig2}
\end{figure*}

\section*{Altermagnetic wallpaper groups}
\setlength{\parindent}{2em}

Exploring beyond monolayer graphene, two-dimensional (2D) honeycomb altermagnets have captured growing research interest, particularly for their unique low-energy physics governed by Dirac-like electronic dispersions. We construct a spin-wallpaper group p6mm in the absence of supercell, respectively flipping the spin on mirror and rotational symmetries related with two opposite-spin mapped to each other\cite{SpinCrystalRotationSymmetry,InversionSymmetry}. The normal subgroup of group p6mm ensures that sublattice symmetries are maintained, where the interplay of mirror $M$ and spin-rotational symmetries stabilizes the spin configurations contributing to the total magnetic moment compensation. Preserving macroscopic net zero magnetization, the symmetry of 2-fold spin rotation maintains the integrity under SO(2) transformation. We expand the unit cell to four sublattices via anti-parallel magnetic moments ($\mu$=$M_{A}-M_{B}$=$M_{C}-M_{D}$) from the hopping of itinerant electrons, where the $\mu$ quantifies the interaction strength between the itinerant electrons and the collinear local moments with staggered magnetization. We emphasize the $XPS_{3}$ (X = Fe, Ni, Mn) family of layered transition-metal phosphorus trichalcogenidesas, which display space group $C2/m$ as exemplary materials for exploring 2D altermagnetism\cite{MagneticGroup,SpinGroup1,SpinGroup2}. These distinct magnetic symmetries and spin configurations reflect the interplay between crystal field anisotropy and exchange interactions within the honeycomb lattice, establishing $XPS_{3}$ compounds as an ideal materials platform for realizing and tuning 2D altermagnetism, where spin compensation and band topology coexist under symmetry protection\cite{MultifunctionalAntiferromagneticMaterials,AltermagneticMetal}.  

\begin{figure}[ht]
\centering
\includegraphics[width=\linewidth]{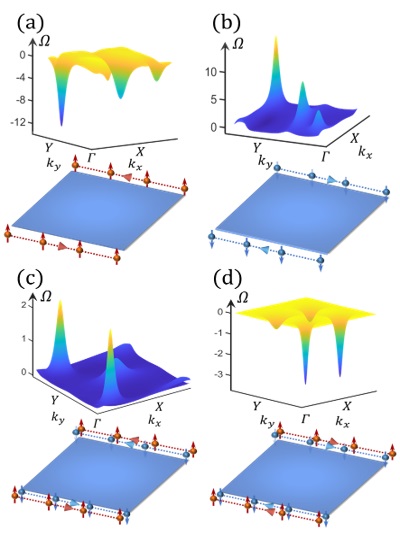}
\caption{The distribution of Berry curvature in the limit of $\nu$$\rightarrow$$\infty$. (a) The directions of spin-up edge currents with $C$ = 1. We use the parameters $t_{2}$=0.6, $\delta_{2}$=1.2, $|\mu|$=0.6, and set the coefficients of mass terms as $m_{1}$=1, $m_{2}$=1.3. (b) The directions of spin-down edge currents with $C$=-1. We use $\delta_{2}$=-1.2 setting the same other parameters as the former. (c) The spin-polarized current with $C$=2. We use the parameters $t_{2}$=0.6, $\delta_{2}$=1.2, $|\mu|$=4 fixing the mass term $m_{2}$=0.3. (d)The spin-polarized current with $C=-2$. We use $\delta_{2}=-1.2$ with the same other parameters as the former.}
\label{fig3}
\end{figure}

We delve into the spectral properties of modified honeycomb lattice, manipulating the position and properties of the Dirac cones through hopping strengths.Spin flip and mirror reflection relate four distinct sublattices as shown in fig. \ref{fig1}, remaining the time-reversal invariant points $\Gamma$ and $M$ in the reduced half BZ. The cell-periodic Hamiltonian is given by

\begin{equation}\label{eq1}
\begin{split}
H_{0}\left(\boldsymbol{k}\right)&=\tau_{0}\otimes2t_{1}\left[\cos\left(k_{1}\right)\cos\left(k_{2}\right)\tau_{x}+\cos\left(k_{1}\right)\sin\left(k_{2}\right)\tau_{y}\right]\\
&-\tau_{z}\otimes2\delta_{1}\left[\sin\left(k_{1}\right)\sin\left(k_{2}\right)\tau_{x}-\sin\left(k_{1}\right)\cos\left(k_{2}\right)\tau_{y}\right]\\
&+\tau_{x}\otimes t_{1}\left[\cos\left(2k_{2}\right)\tau_{x}-\sin\left(2k_{2}\right)\tau_{y}\right]\\
&-\tau_{y}\otimes\delta_{1}\left[\sin\left(2k_{2}\right)\tau_{x}+\cos\left(2k_{2}\right)\tau_{y}\right],\\
\end{split}
\end{equation}

where $\boldsymbol{k}_{1}$=$\frac{\sqrt{3}}{2}\boldsymbol{k}_{x}$, $\boldsymbol{k}_{2}$=$\frac{1}{2}\boldsymbol{k}_{y}$, and $\tau_{i}$ are Pauli matrices acting on the lattice degree of freedom. We reduce the model to a pristine graphene structure with anisotropic nearest-neighbor (NN) hopping $t_{1}$ and $\delta_{1}$, originating from distinct crystalline environments among the four sublattices. A bulk energy gap closing along the boundary of BZ when $\delta_{1}$=0, exhibiting topologically protected linear crossing around the high-symmetry points $(\pm\frac{\pi}{\sqrt{3}}, 0)$ shown in fig\ref{fig1}. (b). The antiparallel orientation of magnetic moments interacts differently with the up and down spin states, which incorporate mirror reflection and the time-reversal symmetry(-$i\sigma_{y}K$). A finite value of $\mu$ lifts the degeneracy of energy bands, except for nodal lines along the high-symmetry directions ($\Gamma-X$, $\Gamma-Y$, $M-X$, and $M-Y$), where additional nodes cross at generic positions between bands sharing the same spin. The band crossings are not restricted to high-symmetry points displaying spin opposition to their mirror-reflected counterparts, which demonstrates the spin-splitting mode of $d_{x^{2}-y^{2}}$ orbitals in the structural renormalization\cite{dWaveAltermagnets}. As $\mu$ increases, they move towards $M$ and eventually merge together to transform the semimetallic state into an insulating state, enabling a variety of intriguing topological phenomena and anisotropic transport properties\cite{SpinHall,AnomalousHall1,AnomalousHall2}. 


\subsection*{Dimerized Model}

The phase diagram with topological invariants outlines the transitions between trivial and topological semimetal phases, capturing the critical localization of robust edge states. We define the homeomorphism $R^{1}\bigcup\infty$$\rightarrow$ $S^{1}$ to illustrate how dimerization generates tunable Dirac cones and induces phase transitions, where the parameter of vertical axis is $\nu$=$\frac{\delta_{1}}{t_{1}}$ with a diameter of $\sqrt{5}$. As shown in Fig.\ref{fig2}. (a), this mapping reveals a fundamental relationship between the dimerization strength and the emergence of massless Dirac fermions, which is protected by the combination of time-reversal symmetry and the bipartite nature of the lattice. The Dirac cones unfold along a high-symmetry path for a set of NN hoppings, eventually merging into a semi-Dirac point characterized at critical values $\nu=0,\pm\frac{\sqrt{5}}{5}, \pm\sqrt{5},\infty$. We highlight the highly anisotropic electronic properties via the convergence of Dirac cones, where direction-dependent behaviors can significantly influence electronic properties such as conductivity, transport responses, and the density of states near the Fermi level. 

We exploit these modifications to preserve the propagation direction of robust edge conduction especially at interfaces, which originates in the two-dimensional Su-Schrieffer-Heeger (SSH) model for polymers\cite{ssh1,ssh2,ssh3}. The emergence of Dirac cones $(\pm\frac{1}{\sqrt{3}}\arccos\frac{\nu^{2}+3}{2\nu^{2}-2}, \frac{\pi}{3})$ at the BZ corners transforms the system back to the semi-metallic phase with the gap size $2\sqrt{1+5\nu^{2}-2\nu\sqrt{5}}$. Based on spinless electrons, the Hamiltonian can be decomposed via a 2$\times$2 block matrix $h_{0}(\boldsymbol{k})$ defined as

\begin{small}
\begin{equation}\label{eq3}
\begin{split}
&t_{1}\left[
\begin{array}{cc}
(1-\nu)e^{i\boldsymbol{a}_{1}}+(1+\nu)e^{-i\boldsymbol{a}_{2}} &  (1+\nu)e^{2i\boldsymbol{k}_{2}}\\
(1-\nu)e^{-2i\boldsymbol{k}_{2}} & (1-\nu)e^{-i\boldsymbol{a}_{2}}+(1+\nu)e^{i\boldsymbol{a}_{1}} \\
\end{array} \right],\\
\end{split}
\end{equation}
\end{small}

where $\boldsymbol{a}_{1}$=$\boldsymbol{k}_{1}-\boldsymbol{k}_{2}$ and $\boldsymbol{a}_{2}$=$\boldsymbol{k}_{1}+\boldsymbol{k}_{2}$. Fixing ${k}_{2}$=0, the $h_{0}\left(\boldsymbol{k}_{1}\right)$ is diagonalized into $h_{0}^{\chi}\left(\boldsymbol{k}_{1}\right)$ = $e^{-ik_{1}}+1$+$\chi\sqrt{\nu^{2}(e^{-ik_{1}}-1)^{2}-\nu^{2}+1}$ through the chiral operator $\chi$. We derive the winding number as a topological invariant protected by chiral symmetry $\chi$=$\tau_{0}\otimes\tau_{z}$.

\begin{equation}\label{eq4}
\begin{split}
w^{\chi}&=\frac{1}{2\pi i}\int d\boldsymbol{k}_{1}\frac{d}{d\boldsymbol{k}_{1}}\log h_{0}^{\chi}\left(\boldsymbol{k}_{1}\right),\\
\end{split}
\end{equation}

where the topological phase transition is quantized by the sign function $sgn(\nu)$, which characterizes a reversal in the chirality of edge mode. The spectrum of nanoribbons has zigzag edges along the $\boldsymbol{k}_{1}$ direction and armchair edges along the $\boldsymbol{k}_{2}$ direction, indicating the existence of propagating states along the transition interface. When $\delta_{1}$=0, the dispersion relation is simplified to four-band analytical structure $E$=$\pm t_{1}\sqrt{\Lambda(\boldsymbol{k})\pm2\sqrt{\lambda(\boldsymbol{k})}}$. 

\begin{small}
\begin{equation}\label{eq2}
\begin{split}
\Lambda(\boldsymbol{k})=&1+4\cos^{2}\left(k_{1}\right)+\nu^{2}\left[1+4\sin^{2}\left(k_{1}\right)\right],\\
\lambda(\boldsymbol{k})=&\left[\cos^{2}\left(3k_{2}\right)+\nu^{2}\sin^{2}\left(3k_{2}\right)\right]\left[4\cos^{2}\left(k_{1}\right)+4\nu^{2}\sin^{2}\left(k_{1}\right)\right]\\
&+\nu^{2}.\\
\end{split}
\end{equation}
\end{small}

where two $\pm$ signs distinguish the particle-hole symmetric branches corresponding to conduction and valence bands. We display the energy spectrum and density of states (DOS) for the edge modes in fig.\ref{fig2}, while spatial distribution of the edge modes is depicted under the open boundary condition with 30 units in the perpendicular direction. Dimerization strengths and specific perturbations can drive the formation and unfolding of the Dirac cones, which are critical for developing next-generation transistors and sensors that leverage high mobility and unique electronic properties\cite{TopologicalInvariants}.


\subsection*{Anomalous Hall Effect}

\begin{figure*}[ht]
\centering
\includegraphics[width=\linewidth]{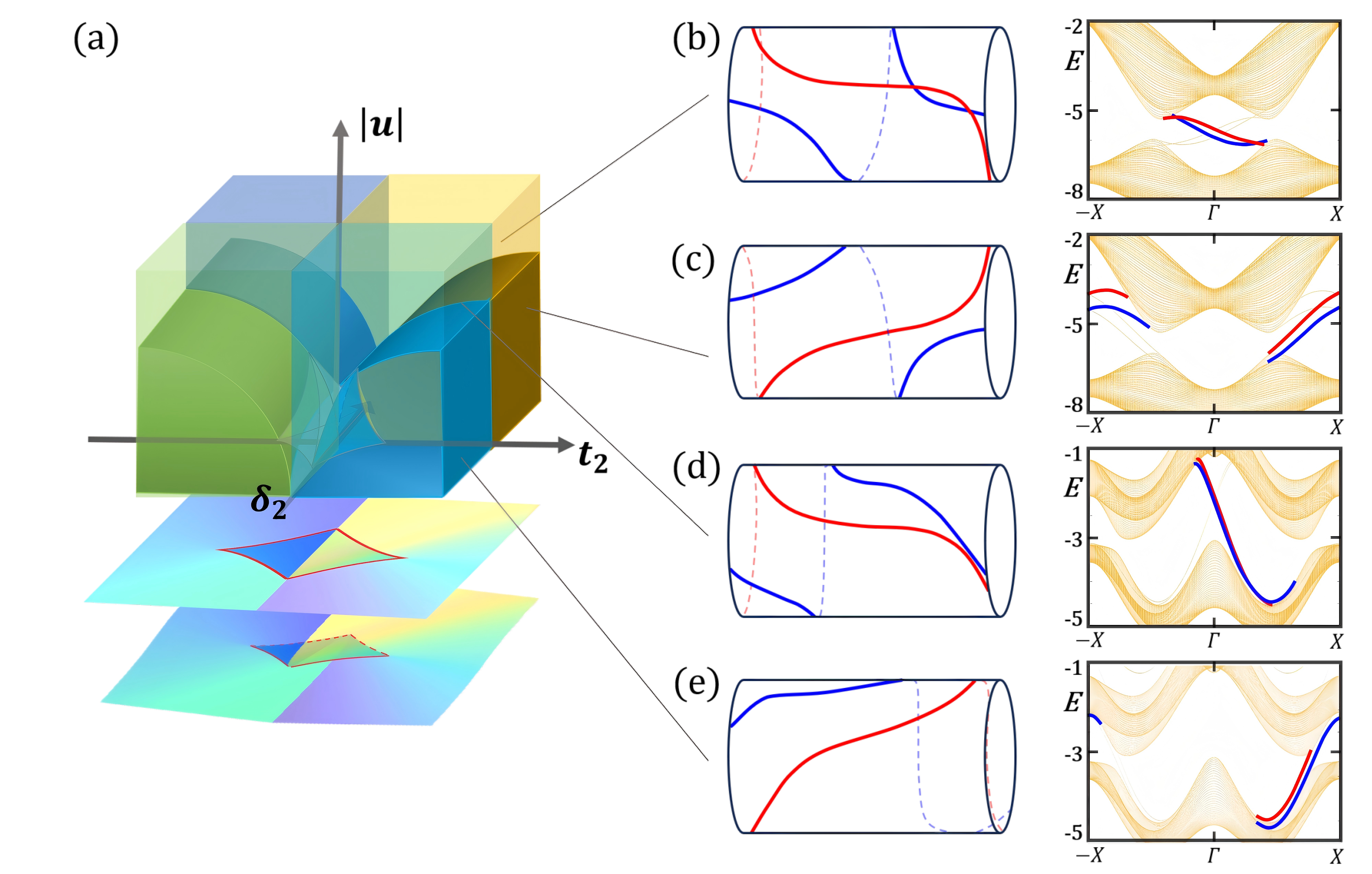}
\caption{(a) The Weyl phase diagram demonstrates the formation of Weyl points, where the phase boundary is related to the singularity class when the map sends regular values to critical values. The number of Weyl points in the vertical axis is determined by the staggered potential $\mu$. (b)-(e) The curved surface separates the creation of weyl points from the gapped phase with $|\mu|$=0.6. The red line shows the chiral propagating states for spin-up, whereas the spin-down counterpart is in blue.}
\label{fig4}
\end{figure*}

Collinear altermagnetic honeycomb lattice with four sublattice arrangements $A, B, C, D$ can generate $\mathcal{T}$-breaking response, notably manifesting as a spontaneous anomalous Hall effect (AHE). In $MnPS_{3}$, there are $N\acute{e}el$-type AFM structure with out-of-plane magnetic moments, lowering the magnetic space group (MSG) to $C2'/m$. The zigzag-AFM configuration parallel to the plane are preserved in $NiPS_{3}$, where the space group and the MSG are transformed to $P_{c}2_{1}/m$ and $P_{s}1$ respectively. Extending the conventional AHE from predominantly associated with ferromagnetic structures, the presence of an out-of-plane component in the anomalous Hall vector facilitates the observation of the AHE in these materials. 

The altermagnetic order is sensitive to the staggered magnetization ($\gamma$=$\tau_{0}\otimes\tau_{z}$) inducing both ferroelectricity and magnetism, where localized spins interact with the itinerant electrons via the Kondo-like coupling $\mu\gamma$ in two decoupled subspaces with well-defined spin. The next-nearest-neighbor (NNN) hoppings are defined as operators $\boldsymbol{S}_{1}$= 2$\delta_{2}\sin\left(\boldsymbol{k}_{1}\right)\cos\left(3\boldsymbol{k}_{2}\right)$$\tau_{y}\otimes\tau_{z}$ and $\boldsymbol{S}_{2}$= -2$\delta_{2}\cos\left(\boldsymbol{k}_{1}\right)\sin\left(3\boldsymbol{k}_{2}\right)$$\tau_{y}\otimes\tau_{0}$, where another term related to the NNN hopping $t_{2}$ is $\boldsymbol{S}_{3}$=4$t_{2}\cos\left(\boldsymbol{k}_{1}\right)\cos\left(3\boldsymbol{k}_{2}\right)$$\tau_{x}\otimes\tau_{0}$. Thus, we can construct altermagnetic model from the $\theta$=$\pi$ singularity of $S^{1}$ in fig. \ref{fig2}.

\begin{equation}\label{eq5}
\begin{split}
H\left(\boldsymbol{k}\right)&=\sigma_{0}\otimes H_{1}\left(\boldsymbol{k}\right)+\mu\sigma_{z}\otimes\gamma ,\\
H_{1}\left(\boldsymbol{k}\right)&=H_{0}\left(\boldsymbol{k}\right)+\boldsymbol{S}_{1}+\boldsymbol{S}_{2}+\boldsymbol{S}_{3},\\
\end{split}
\end{equation}

where $\sigma_{i}$ are Pauli matrices operating on the spin group. The quadratic energy spectrum becomes subject to the altermagnetic potentials that lift the spin-degeneracy. We calculate the energy spectrum for the $\mu\gamma$ combined with staggered NNN hoppings as $E^{\chi}$=$\chi\mu\pm$ $2\sqrt{4t_{2}^{2}\cos^{2}(k_{1})\cos^{2}(3k_{2})+\delta_{2}^{2}\sin^{2}(k_{1}+3\chi k_{2})}$. 

Without considering the SOC, the dispersion relation approaches a linear behavior, where two spin-degenerate points can form a fourfold degeneracy near the Fermi level with additional mirror symmetry protection. The bands of opposite spins hold the topologically protected 4-fold Dirac crossings on the $MX$ line owing to the mirror operation not swapping the sublattices. The mass term $M_{\boldsymbol{k}}$=$m_{1}\cos(2\boldsymbol{k}_{1})$+$m_{2}\sin(2\boldsymbol{k}_{1})$+$m_{3}\sin(2\boldsymbol{k}_{2})$ is also introduced by multiplying the general staggered hoppings with the lattice space operator $\tau_{z}\otimes\tau_{0}$. Thus, the topology of crossing is characterized by the Berry Phase $\Upsilon^{\alpha(\beta)}_{\pm}$.

\begin{equation}
\begin{split}
\Upsilon^{\alpha(\beta)}_{\pm}&=\pm\frac{1}{4\pi}\int_{BZ}\frac{\boldsymbol{d}^{\alpha(\beta)}_{\boldsymbol{q}} \cdot[\partial_{q_{x}}\boldsymbol{d}^{\alpha(\beta)}_{\boldsymbol{q}} \times\partial_{q_{y}}\boldsymbol{d}^{\alpha(\beta)}_{\boldsymbol{q}}]}{|\boldsymbol{d}^{\alpha(\beta)}_{\boldsymbol{q}}|^{3}},\\
&=\pm\frac{1}{4\pi}\int_{BZ}\frac{M_{\boldsymbol{k}}v_{x}v_{y}}{2\sqrt[3]{v_{x}q^{2}_{y}+v_{y}q^{2}_{x}+M^{2}}} =\pm\frac{sgn(M_{\boldsymbol{k}})}{2},\\
\end{split}
\end{equation} 

where $\boldsymbol{d}^{\alpha(\beta)}_{\boldsymbol{q}}$=$v^{\alpha(\beta)}_{y}$$\boldsymbol{q}_{x}$+ $v^{\alpha(\beta)}_{x}$$\boldsymbol{q}_{y}$+$\frac{M_{\boldsymbol{k}}}{2}$$\boldsymbol{q}_{z}$. The distribution of Berry curvature contributes to the total Hall current as $j$=$\frac{\hbar}{2e}$$(j^{\alpha}$+$j^{\beta})$=$\frac{e^{2}}{\hbar}$$\sum^{\alpha, \beta}sgn(M_{\boldsymbol{k}})$\cite{SpinHall}. The tilted Dirac points exhibit robust stability protected by time-reversal and spin-inversion symmetries, while the mirror symmetry acts as a stabilizing operator for Dirac points, remaining degenerate along the mirror-symmetric lines.

\subsection*{Weyl Phase Digram}

We study the splitting of energy bands that contribute to the formation of linear crossings protected by parity with spin and lattice indices independently. Parity operator $(x, y) \rightarrow (-x, -y)$ preserve both spin and sublattice degrees of freedom in the absence of SOC, mapping a Weyl point onto its counterpart with an odd (even) parity eigenvalue. Similar to the topological protection of Weyl points, the crossings are protected by the combined action of time-reversal symmetry and mirror symmetry\cite{TimeReversalSymmetry,WeylNodalLoops}. The Weyl crossings are isolated in momentum space evolving along the high-symmetry directions\cite{WeylPoints,WeylFermions}. The mirror symmetry $\tau_{0}\otimes\tau_{x}$ is combined with the time-reversal symmetry ensuring the symmetry compensation in altermagnets\cite{CrystalSymmetry}. Despite the spin splitting, the zero net magnetic moment is attributed to the combined symmetry that acts non-trivially on spin and cell\cite{ClassificationSymmetry}. The Weyl crossings between bands of the same spin but different sublattices undergo a topological transition to trivial bands upon increasing the magnitude of the magnetization in altermagnets\cite{WeylSemimetal1,WeylSemimetal2,mapusuo2}. Acting as monopoles or anti-monopoles of Berry curvature, the gapped Weyl crossings enable the valley-dependent Hall effect, chiral magnetic effects, and other exotic transport phenomena\cite{SpinMomentumLockedTransport}.

We depict the relevant Wannier center flows by non-Abelian Berry phase in fig.\ref{fig4}, corresponding to the integration of Berry curvature with fully decoupled spin subspaces\cite{BerryPhase1,BerryPhase2}. As a critical point in fig. \ref{fig2}, the $t_1$ term opens the altermagnetic nodal line generating the spin-polarized Dirac cones with $\delta_{1}$=0. We map the effective Hamiltonian into the subspace via the operator $\boldsymbol{P}$=$|\Psi^{\alpha}(\boldsymbol{k})\rangle$ $\langle\Psi^{\alpha}(\boldsymbol{k})|$$\bigoplus$$|\Psi^{\beta}(\boldsymbol{k})\rangle$ $\langle\Psi^{\beta}(\boldsymbol{k})|$. 

\begin{equation}\label{eq6}
\begin{split}
H_{q^{\alpha}}&=\epsilon^{\alpha}_{q}\sigma_{0}+v^{\alpha}_{y}q^{\alpha}_x\sigma_{y}\pm v^{\alpha}_{x}q^{\alpha}_y\sigma_{x},\\
H_{q^{\beta}}&=\epsilon^{\beta}_{q}\sigma_{0}+v^{\beta}_{y}q^{\beta}_x\sigma_{y}\pm v^{\beta}_{x}q^{\beta}_y\sigma_{x},\\
\end{split}
\end{equation}

where $\epsilon^{\alpha}_{\boldsymbol{q}}$=$\epsilon^{\beta}_{\boldsymbol{q}}$=0 are half filled Fermi energy. The anisotropic velocities for spin-up subspace are $v^{\alpha}_{x}$ and $v^{\alpha}_{y}$, where the counterparts for spin-down are represented by ${\beta}$. The presence of $t_{1}$ reshapes the subspace algebraic structure via $\xi$=$|\frac{R^{2}}{2\delta_{2}^{2}}-1|$, $R^{2}$=$t_{1}^{2}+\mu^{2}$. We derive the anisotropic velocities in the vicinity of pair-wise crossings $(\pm\frac{\pi}{\sqrt{3}}$,$\pm\frac{\pi-\arccos\sqrt{\xi}}{3})$.

\begin{equation}
\begin{split}
v^{\alpha}_{y}&=v^{\beta}_{y}=\delta_{2}\sqrt{3\xi},\\
v^{\alpha\beta}_{x}&=\sqrt{|9\delta_{2}^{2}-9R^{2}+\chi18R^{2}\cos^{4}\frac{\sqrt{1-\xi}}{3}|}.\\
\end{split}
\end{equation} 

The Weyl phase diagram is formed by $t_{2}$ and $\delta_{2}$ in the configurational space fixing $t_{1}$=0 and $\delta_{1}$=1. The Wilson loop approach describes the evolution of Wannier function center based on the $U(2N)$ Berry connection $A^{\alpha\beta}$= $\langle\psi^{\alpha}(\boldsymbol{k})|\psi^{\beta}(\boldsymbol{k}+\delta_{\boldsymbol{k}})\rangle$. We project the $H_{0}\left(\boldsymbol{k}\right)$ into the Bloch occupied states via the position operator 
$\sum_{\boldsymbol{k}}$$\sum^{N_{F}}_{\alpha, \beta=1}$$A^{\alpha\beta}$$|\Psi^{\beta}(\boldsymbol{k}+\delta_{\boldsymbol{k}})\rangle$ $\langle\Psi^{\alpha}(\boldsymbol{k})|$. Typical phases are illustrated in fig.\ref{fig4}, which are modified by varying the staggered magnetization as the control parameter. We bridge a connection between the mappings of manifolds $M^{2}$$\rightarrow$$R^{2}$, where $R^{2}$ is two-dimensional Euclidean space. The emergence of pairwise Weyl points creates a cusp point from the map self-intersection.

The spin degeneracy of the edge modes is lifted allowing for unique chiral propagating states under the $Z_2$ topological classification\cite{ZakPhase}.  We theoretically calculate the quantum Hall effects via the controlled fabrication of the valley degree of freedom in altermagnetic band structure in fig.\ref{fig4}. The formation of Weyl points enables nontrivial topology for spin splitting band structure, acting as monopoles or anti-monopoles of Berry curvature determined by their chirality as shown in fig.\ref{fig3}. We depict the zigzag edges of altermagnetic ribbon, where the magnetic moments on adjacent sites are antiparallel. Induced by the antiparallel orientation of magnetic moments, the gap opening at the Weyl points illustrates chiral transport responses rather than helical spin textures\cite{SpontaneousHall1,SpontaneousHall2}. We suggest the injection of high-energy photons can induce non-trivial topological states in our altermagnetic model, where the radiative recombination of spin-polarized carriers exhibits a net circular polarization\cite{PNJunction1}. These states can also be realized in non-Hermitian systems through the modulation of gain and loss, both of which are guided by mirror symmetry and $PT$ symmetry\cite{GrapheneHeterojunctions}.


\section*{Possible experimental realizations}

The application of an out-of-plane electric field can modulate the magnetic interactions to more complex magnetic orderings\cite{MnTe1,MnTe3}, while the unique interplay of lattice symmetries suggests $MnP(S,Se)_{3}$ might be functionalized into a strong altermagnet from a 2D antiferromagnet\cite{MnPSeFeSe}. The structural phase transition induced by chalcogen substitution modifies the exchange pathways, potentially creating the spin-split electronic bands characteristic of altermagnetic materials while preserving the compensated antiparallel spin arrangement. The particular $C_{2}$ rotational symmetry of the Mn cation sublattice, combined with the ligand-field splitting induced by chalcogen substitution, creates the necessary conditions for alternating spin-polarized bands in both real and reciprocal space. Through precise strain engineering, this structural configuration may enable $N\acute{e}el$ temperatures surpassing room temperature, which would be significant for practical applications. The time-reversal symmetry breaking in monolayer FeSe is also not simply a manifestation of ferromagnetism but rather an indication of altermagnetism\cite{La2O3Mn2Se2,gifwave}. These observations offer strong evidence for this unconventional magnetic ordering phenomenon. Combined with the Kerr effect, the polarized photons interacting with the staggered magnetic moments can regulate topological responses in advanced optoelectronic devices\cite{PNJunction2,PNJunction3}. 

Ultracold atoms in optical lattices offer a highly controllable platform for studying altermagnetic topology\cite{UltracoldAtoms}. The precise manipulation of atomic interactions and lattice geometries enables the emulation of exotic quantum phases, including those exhibiting altermagnetic order with non-trivial band topology. By tuning the lattice parameters, we can engineer our model into different topological regimes via artificial mechanical lattices, allowing for the observation of the robust transport for chiral edge modes. Additionally, in ultracold atomic gases, the implementation of synthetic magnetic fields provides a controlled platform for engineering diverse topological phases and observing valley-dependent Hall effects\cite{ValleyTransport1,ValleyTransport2}. Topolectrical circuits hold promise to manipulate the valley polarization through topological boundary resonances\cite{QuantumCircuits1,QuantumCircuits2}, enabling selective excitation of valley-polarized states through appropriate impedance modulation. We consider a honeycomb lattice structure with engineered parameters to achieve bond dimerization and onsite potentials, where Weyl circuits exhibit tunable edge states governed by their intrinsic topology. The valley-polarized modes demonstrate valley-contrasting transport properties, while the valley index provides an additional degree of freedom for information encoding\cite{QuantumCircuits3,QuantumCircuits4,QuantumCircuits5}. Topolectrical circuits exploit the inherent degeneracy to engineer robust states that are immune to backscattering and disorder, enabling the realization of the valleytronic devices and the application of topological quantum computing\cite{QuantumCircuits6,QuantumCircuits7}.

\section*{Declarations}
Meng-Han Zhang wrote the main manuscript text and Xuan Guo prepared figures 1-4. All authors reviewed the manuscript. We declare that the authors have no competing interests as defined by Nature Portfolio, or other interests that might be perceived to influence the results and/or discussion reported in this paper. We thank Zi-Jian Xiong for helpful discussions. 

\section*{Funding Declaration
}
This project is supported by NKRDPC-2022YFA1402802, NSFC-92165204, NSFC-12494590, Leading Talent Program of Guangdong Special Projects (201626003), Guangdong Provincial Key Laboratory of Magnetoelectric Physics and Devices (No. 2022B1212010008), Research Center for Magnetoelectric Physics of Guangdong Province(2024B0303390001), and Guangdong Provincial Quantum Science Strategic Initiative (GDZX2401010).

\bibliography{alterMTBFinal}

\end{document}